\documentclass[prc,aps,twocolumn,superscriptaddress,amsmath,amssymb,floatfix]{revtex4-2}
\usepackage[colorlinks=true,citecolor=blue,filecolor=blue,linkcolor=blue,urlcolor=blue]{hyperref}
\usepackage{orcidlink}

\usepackage{float}
\usepackage{graphicx}% Include figure files
\usepackage{dcolumn}% Align table columns on decimal point
\usepackage{bm}% bold math
\usepackage{siunitx}
\usepackage{multirow}%
\usepackage{amsmath,amssymb,amsfonts}%
\usepackage{amsthm}%
\usepackage{mathrsfs}%
\usepackage[title]{appendix}%
\usepackage{xcolor}%
\usepackage{textcomp}%
\usepackage{booktabs}%
\usepackage{algorithm}%
\usepackage{algorithmicx}%
\usepackage{algpseudocode}%
\usepackage{listings}%
\usepackage[left,mathlines]{lineno}%
\begin{document}

\title{Enhancing Neutrinoless Double-Beta Decay Sensitivity of Liquid-Xenon Time Projection Chamber with Augmented Convolutional Neural Network}

% Autogenerated on 05 Mar, 2026 at 12:49:30
% Generated with make-authorlist.py

% -> SNIP
% \documentclass[prl,aps,twocolumn,superscriptaddress,amsmath,amssymb,floatfix]{revtex4}
% \usepackage[colorlinks=true,citecolor=blue,filecolor=blue,linkcolor=blue,urlcolor=blue,pdftex]{hyperref}
% \usepackage{orcidlink} % For ORCID links in author list

% \usepackage{currfile} % To be able to list current file name in title
% <- SNIP

% You could put the following into an 'affiliation.tex' file and 
% \include it in your main text

\newcommand{\bologna}{\affiliation{Department of Physics and Astronomy, University of Bologna and INFN-Bologna, 40126 Bologna, Italy}}
\newcommand{\chicago}{\affiliation{Department of Physics, Enrico Fermi Institute \& Kavli Institute for Cosmological Physics, University of Chicago, Chicago, IL 60637, USA}}
\newcommand{\coimbra}{\affiliation{LIBPhys, Department of Physics, University of Coimbra, 3004-516 Coimbra, Portugal}}
\newcommand{\columbia}{\affiliation{Physics Department, Columbia University, New York, NY 10027, USA}}
\newcommand{\lngs}{\affiliation{INFN-Laboratori Nazionali del Gran Sasso and Gran Sasso Science Institute, 67100 L'Aquila, Italy}}
\newcommand{\mainz}{\affiliation{Institut f\"ur Physik \& Exzellenzcluster PRISMA$^{+}$, Johannes Gutenberg-Universit\"at Mainz, 55099 Mainz, Germany}}
\newcommand{\mpik}{\affiliation{Max-Planck-Institut f\"ur Kernphysik, 69117 Heidelberg, Germany}}
\newcommand{\munster}{\affiliation{Institut f\"ur Kernphysik, University of M\"unster, 48149 M\"unster, Germany}}
\newcommand{\nikhef}{\affiliation{Nikhef and the University of Amsterdam, Science Park, 1098XG Amsterdam, Netherlands}}
\newcommand{\nyuad}{\affiliation{New York University Abu Dhabi - Center for Astro, Particle and Planetary Physics, Abu Dhabi, United Arab Emirates}}
\newcommand{\purdue}{\affiliation{Department of Physics and Astronomy, Purdue University, West Lafayette, IN 47907, USA}}
\newcommand{\rice}{\affiliation{Department of Physics and Astronomy, Rice University, Houston, TX 77005, USA}}
\newcommand{\stockholm}{\affiliation{Oskar Klein Centre, Department of Physics, Stockholm University, AlbaNova, Stockholm SE-10691, Sweden}}
\newcommand{\subatech}{\affiliation{SUBATECH, IMT Atlantique, CNRS/IN2P3, Nantes Universit\'e, Nantes 44307, France}}
\newcommand{\torino}{\affiliation{INAF-Astrophysical Observatory of Torino, Department of Physics, University  of  Torino and  INFN-Torino,  10125  Torino,  Italy}}
\newcommand{\ucsd}{\affiliation{Department of Physics, University of California San Diego, La Jolla, CA 92093, USA}}
\newcommand{\wis}{\affiliation{Department of Particle Physics and Astrophysics, Weizmann Institute of Science, Rehovot 7610001, Israel}}
\newcommand{\zurich}{\affiliation{Physik-Institut, University of Z\"urich, 8057  Z\"urich, Switzerland}}
\newcommand{\paris}{\affiliation{LPNHE, Sorbonne Universit\'{e}, CNRS/IN2P3, 75005 Paris, France}}
\newcommand{\freiburg}{\affiliation{Physikalisches Institut, Universit\"at Freiburg, 79104 Freiburg, Germany}}
\newcommand{\napels}{\affiliation{Department of Physics ``Ettore Pancini'', University of Napoli and INFN-Napoli, 80126 Napoli, Italy}}
\newcommand{\nagoya}{\affiliation{Kobayashi-Maskawa Institute for the Origin of Particles and the Universe, and Institute for Space-Earth Environmental Research, Nagoya University, Furo-cho, Chikusa-ku, Nagoya, Aichi 464-8602, Japan}}
\newcommand{\laquila}{\affiliation{Department of Physics and Chemistry, University of L'Aquila, 67100 L'Aquila, Italy}}
\newcommand{\tokyo}{\affiliation{Kamioka Observatory, Institute for Cosmic Ray Research, and Kavli Institute for the Physics and Mathematics of the Universe (WPI), University of Tokyo, Higashi-Mozumi, Kamioka, Hida, Gifu 506-1205, Japan}}
\newcommand{\kobe}{\affiliation{Department of Physics, Kobe University, Kobe, Hyogo 657-8501, Japan}}
\newcommand{\kit}{\affiliation{Institute for Astroparticle Physics, Karlsruhe Institute of Technology, 76021 Karlsruhe, Germany}}
\newcommand{\tsinghua}{\affiliation{Department of Physics \& Center for High Energy Physics, Tsinghua University, Beijing 100084, P.R. China}}
\newcommand{\ferrara}{\affiliation{INFN-Ferrara and Dip. di Fisica e Scienze della Terra, Universit\`a di Ferrara, 44122 Ferrara, Italy}}
\newcommand{\groningen}{\affiliation{Nikhef and the University of Groningen, Van Swinderen Institute, 9747AG Groningen, Netherlands}}
\newcommand{\westlake}{\affiliation{Department of Physics, School of Science, Westlake University, Hangzhou 310030, P.R. China}}
\newcommand{\shenzhen}{\affiliation{School of Science and Engineering, The Chinese University of Hong Kong (Shenzhen), Shenzhen, Guangdong, 518172, P.R. China}}
\newcommand{\coimbrapoli}{\affiliation{Coimbra Polytechnic - ISEC, 3030-199 Coimbra, Portugal}}
\newcommand{\uniheidelberg}{\affiliation{Physikalisches Institut, Universit\"at Heidelberg, Heidelberg, Germany}}
\newcommand{\roma}{\affiliation{INFN-Roma Tre, 00146 Roma, Italy}}
\newcommand{\bucknell}{\affiliation{Department of Physics \& Astronomy, Bucknell University, Lewisburg, PA, USA}}

% End of AFFILIATIONS

% \begin{document}

% CHANGE TO TITLE
% \title{XENON Authors \currfilename}

% You could put the following into an 'authors.tex' file and 
% \include it in your main text

\author{E.~Aprile\,\orcidlink{0000-0001-6595-7098}}\columbia
\author{J.~Aalbers\,\orcidlink{0000-0003-0030-0030}}\groningen
\author{K.~Abe\,\orcidlink{0009-0000-9620-788X}}\tokyo
\author{M.~Adrover\,\orcidlink{0123-4567-8901-2345}}\zurich
\author{S.~Ahmed Maouloud\,\orcidlink{0000-0002-0844-4576}}\paris
\author{L.~Althueser\,\orcidlink{0000-0002-5468-4298}}\munster
\author{B.~Andrieu\,\orcidlink{0009-0002-6485-4163}}\paris
\author{E.~Angelino\,\orcidlink{0000-0002-6695-4355}}\lngs
\author{D.~Ant\'on~Martin\,\orcidlink{0000-0001-7725-5552}}\chicago
\author{S.~R.~Armbruster\,\orcidlink{0009-0009-6440-1210}}\mpik
% ...
\collaboration{XENON Collaboration}\email[]{xenon@lngs.infn.it}\noaffiliation

%
% End of AUTHORS

% \date{\today} 

% CHANGE
% \begin{abstract}
% \end{abstract}

% <--- SNIP 
% \maketitle

% \end{document}

\date{\today}% It is always \today, today,

\begin{abstract}
Dual-phase time projection chamber (TPC) that employs a multi-ton-scale liquid xenon (LXe) target mass is a pioneering detector technology to search for dark matter. Beyond its advantage in dark matter direct detection efforts, the natural xenon target allows it to search for the neutrinoless double-beta decay ($0\nu\beta\beta$) process, which would violate lepton number conservation and indicate that neutrinos are Majorana particles. However, such $0\nu\beta\beta$ searches have been limited by gamma-ray backgrounds originating from the detector materials. In this work, we designed an augmented convolutional neural network (A-CNN) model to extract additional event-topology information from detector data. Using simulation and calibration data from XENONnT, a leading LXe TPC experiment, our model achieved over 60\% background rejection while maintaining 90\% signal acceptance. This rejection power improves XENONnT's projected sensitivity of the $^{136}$Xe $0\nu\beta\beta$ search by about 40\%. The implementation of A-CNN in the data analysis of future liquid xenon observatories, such as XLZD, will further enhance their sensitivities for $0\nu\beta\beta$ with $^{136}$Xe.
\end{abstract}

\maketitle

%\tableofcontents

\definecolor{BLUE}{rgb}{0,0,1}
\definecolor{RED}{rgb}{1,0,0}

\section{Introduction}\label{sec1}

\subsection{Neutrinoless Double-Beta Decay}\label{subsec11}

% \begin{itemize}
%   \item Rare event searches require highly customized detectors and low background, which requires great analysis effort.
%   \item Machine learning is a powerful tool to analyze the sparse data and decode the information hidden in low-level data.
%   \item Neutrinoless double-beta decay is one of the most important topics in modern physics that once observed could prove its Majorana nature and provide insights into matter-antimatter symmetry.
%   \item The paper is structured as follows, ..., including highlights.
% \end{itemize}
Rare-event search experiments provide a unique way to understand the universe. As one of the primary frontiers in rare-event searches, neutrinoless double beta decay ($0\nu\beta\beta$) is a hypothetical process~\cite{Furry:1939qr}, which if found, will prove that neutrinos are their own antiparticles. This theorized Majorana nature is also a key ingredient in some mechanisms, such as leptogenesis, that could explain the observed matter–antimatter asymmetry in our universe~\cite{Fukugita:1986hr}. $0\nu\beta\beta$ is difficult to observe because of its ultra-long decay half-life, in that even a tonne-scale experiment might not be able to detect one signal after years of data taking. This places stringent requirements on the identification and suppression of all background events to maximize the possibility of $0\nu\beta\beta$ discovery.

Liquid xenon time projection chamber (LXe TPC) detectors are leading technologies in rare-event searches. They can be roughly divided into two categories: the single-phase TPC and the dual-phase TPC. The single-phase TPC has been used by the concluded EXO-200 experiment~\cite{EXO-200:2019rkq} and proposed nEXO experiment~\cite{nEXO:2021ujk} to search for $0\nu\beta\beta$. The EXO-200 experiment has reported a lower limit on the $^{136}$Xe $0\nu\beta\beta$ half-life of $3.5\times 10^{25}$ yr at 90\% CL~\cite{EXO-200:2019rkq}, while the proposed nEXO experiment has a projected sensitivity above $10^{28}$ yr~\cite{nEXO:2021ujk}.

On the other hand, dual-phase TPC technology can be used to search for both WIMP dark matter~\cite{XENON:2025vwd} and $0\nu\beta\beta$. The XENON1T~\cite{XENON:2022evz} and PandaX-4T~\cite{pandax0vbb} experiments leveraging this technology have reported their $0\nu\beta\beta$ decay search limits in the past. For example, the concluded XENON1T experiment obtained a half-life lower limit for $^{136}$Xe $0\nu\beta\beta$ of $1.2\times10^{24}$ yr at 90\% CL. In XENONnT, its upgraded version, the projected median lower limit for $^{136}$Xe $T_{1/2}^{0\nu\beta\beta}$ was $2.1\times10^{25}$ yr at 90\% CL~\cite{XENON:2022evz}, due to the larger mass and lower background level. Meanwhile, non-TPC $0\nu\beta\beta$-dedicated experiments like KamLAND-Zen also reported a lower limit on the $^{136}$Xe $0\nu\beta\beta$ half-life of $3.8\times 10^{26}$ yr at 90\% CL~\cite{KamLAND-Zen:2024eml}, benefited from its large target and low background, which currently represents the most stringent limit to date. The xenon-based dark matter experiments so far reported worse limits than those of dedicated $0\nu\beta\beta$ experiments like EXO-200 or KamLAND-Zen. One major reason for this drawback is the material background: while dark-matter detectors focus on developing novel technologies to achieve ultra-low-energy backgrounds below 100 keV, $0\nu\beta\beta$ of $^{136}$Xe has a Q-value of 2.458 MeV, which is well beyond the low-background energy range. This means radioactive background from detector materials can produce events within the $0\nu\beta\beta$ region of interest, which significantly hinders the sensitivity to $0\nu\beta\beta$.

Machine learning (ML) provides a cost-effective solution to this challenge without relying on expensive detector upgrades or background removal campaigns. In this work, we introduce a novel machine learning algorithm that uses the charge signals to understand event topologies and is capable of removing the dominant background for the $0\nu\beta\beta$ decay search in LXe TPC detectors. Using XENONnT's simulation and calibration data as a demonstration, our sensitivity study shows that the successful implementation of this algorithm can boost the sensitivity of dual-phase LXe TPCs to $0\nu\beta\beta$ $T_{1/2}^{0\nu\beta\beta}$ by about 40\% without additional hardware investment. The paper is structured as follows. Section~\ref{sec12} introduces LXe TPC and its data type, Section~\ref{sec2} explains the main background sources and expected sensitivity of the $0\nu\beta\beta$ search at XENONnT. Section~\ref{sec3} illustrates the Monte-Carlo simulation for the training data preparation, the augmentation technique, and the A-CNN (convolutional neural network with data augmentation) network design. Section~\ref{sec4} outlines the training results, data-MC matching, and the interpretability study of the A-CNN model. We highlight A-CNN's ability to interpret its decision via the saliency map, thus unveiling that the underlying physics is the source of the discriminating power. Section~\ref{sec5} introduces the application of the A-CNN to improve the expected sensitivity of the $0\nu\beta\beta$ search with XENONnT.

\subsection{Dual-Phase LXe TPC and Its Data}\label{sec12}

\begin{figure*}[htbp]
\centering
\includegraphics[width=0.95\textwidth]{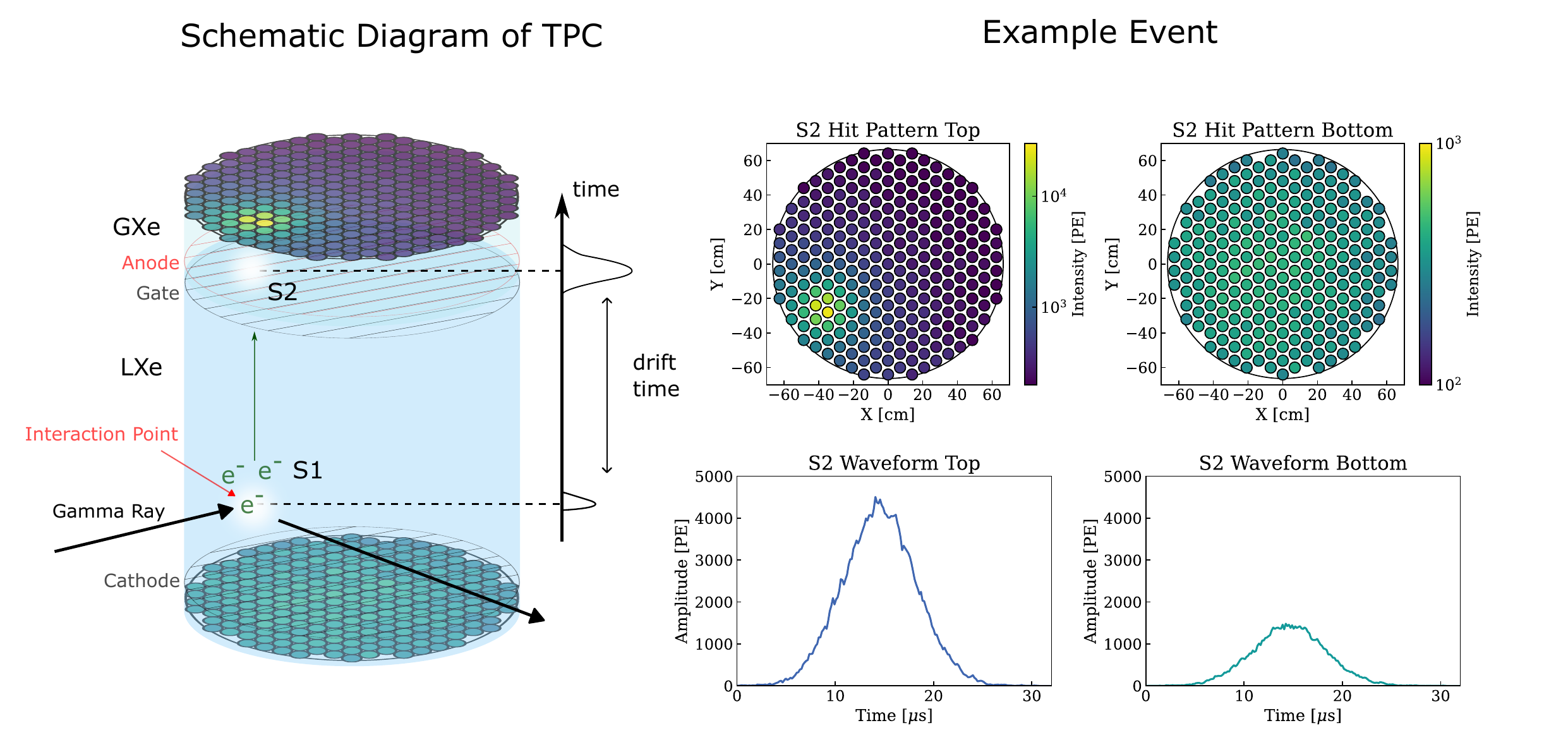}
\caption{Left: TPC schematic with illustration of a gamma event. Right: top and bottom PMT hit patterns and waveforms.}\label{figure-xenonschematic}
\end{figure*}

This section provides a general introduction to dual-phase LXe TPC detectors and the data they produce, using XENONnT, a world-leading dark matter experiment, as an example. The XENONnT experiment~\cite{XENON:2024wpa} is operated by the XENON collaboration at the INFN Laboratori Nazionali del Gran Sasso (LNGS) underground laboratory. Figure~\ref{figure-xenonschematic} (Left) shows the schematic diagram of the dual-phase LXe TPC. XENONnT has a new, larger TPC compared to XENON1T, featuring a sensitive LXe mass of 5.9 t, viewed by 494 PMTs distributed across the top and bottom arrays. When particles deposit energy in the detector, the prompt scintillation photons are detected as S1 signals, while the ionization electrons drift toward the top of the TPC under the electric field and generate electroluminescence near the liquid–gas interface, which is then detected by the PMTs as S2 signals. The collected data contains time and space information, where the hit pattern of the PMTs is used to infer the X-Y position of the event, and the time difference or drift time between the prompt S1 signal and delayed S2 signal is used for inferring the depth information of the event. A total of 8.5 t LXe is continuously purified by a new combined liquid- and gas-phase purification system~\cite{Plante:2022khm}. Together with a high-flow radon distillation system~\cite{Murra:2022mlr}, a careful selection of detector construction materials~\cite{XENON:2021mrg} and a specialized assembly procedure, this has led to the lowest electron recoil (ER) background level ever achieved below 100 keV~\cite{XENON:2022ltv}.

Figure \ref{figure-xenonschematic} (Right) shows example hit patterns and the summed waveforms of the top and bottom PMT arrays from a $\sim$MeV event. The recorded signals are processed using the custom-developed open-source software package Strax and Straxen~\cite{straxen}. Each PMT signal is scanned for PMT “hits” above a threshold, and hits found in the TPC channels are clustered and classified into S1, S2 or “unclassified” peaks based on pulse shape and PMT hit pattern. Although the original per-channel PMT waveforms are in spatiotemporal form, we summed them over the channel dimension into 1D time series (waveforms), to minimize computational demands. More details about the signal reconstruction can be found in~\cite{XENONCollaboration:2024bil}.

%  At least three PMTs must contribute to an S1 within ±50 ns around the center of the integrated peak waveform. Events are built in time intervals between 2.45 ms before and 0.25 ms after S2s, and overlapping events are merged. The event's S2 is required to be greater than 100 PE, and have fewer than eight other peaks larger than half of the S2 peak area within ±10 ms \cite{XENON:2023cxc}, to avoid multi-scatter events and busy times when peaks might be misreconstructed with accidentally coincident S1s and S2s that are actually unrelated.

In dual-phase TPCs, S1 signals are typically fast on the order of tens of nanoseconds, providing limited information for distinguishing between single-site (SS) and multi-site (MS) events. In contrast, S2 signals originate near the liquid–gas interface at the top of the detector, making the top PMT array more sensitive to their features, whereas photons traveling to the bottom array undergo longer propagation paths that result in smeared signals. Therefore, in this work, we only use S2 signals seen by the top PMT array (S2$_{\text{top}}$) for the development of the ML model, to preserve most of the interesting part of the event and cut unnecessary computing resource requirements.

In this work, the goal is to provide an illustrative example of developing ML models for LXe TPCs that can capture more information from the low-level detector data and improve signal and background classification, which is key in rare-event searches like $0\nu\beta\beta$. The resulting model is an augmented convolutional neural network (A-CNN)~\cite{oshea2015introduction, 2017arXiv171204621P} which is more robust and adaptable in physics searches compared to conventional CNN. The enhanced A-CNN algorithm shows strong discrimination power in event topologies, which can be applied to the experimental data analysis of dual-phase LXe TPCs for the $0\nu\beta\beta$ search.

\section{Background and \texorpdfstring{$0\nu\beta\beta$}{0nuBB} Sensitivity}\label{sec2}

The major factors that affect the experimental sensitivity to $0\nu\beta\beta$ include the exposure and the background. The typical half-life sensitivity is described by~\cite{DolinskiPoonRodejohann:2019}:
\begin{equation}
\label{eq:halflife-scaling}
T_{1/2}^{0\nu\beta\beta} \propto \begin{cases} 
a M \epsilon t & \text{background free} \\
a \epsilon \sqrt{\frac{M t}{B \Delta E}} & \text{with background,}
\end{cases}
\end{equation}
where $T_{1/2}^{0\nu\beta\beta}$ is the $0\nu\beta\beta$ half-life that can be measured or constrained by the experiment; $M$ is the total target mass and $a$ is the isotopic abundance, representing the percentage of $0\nu\beta\beta$ candidate isotopes within the total isotopic mass; $\epsilon$ is the detection efficiency of the signal in the region of interest; $t$ is the measurement time; $\Delta E$ is the detector energy resolution and $B$ is the background index, or the number of background counts normalized to the width of the energy region of interest (ROI), isotopic mass $aM$, and measurement time $t$. Therefore, $B$ usually has the unit of (keV$\cdot$kg$\cdot$yr)$^{-1}$. This equation shows that in the background-free case, $T_{1/2}^{0\nu\beta\beta}$ is linearly proportional to the detector mass ($M$). However, if backgrounds exist, the sensitivity depends on the square root of mass and time, and is inversely proportional to the background index $B$ and energy resolution $\Delta E$. 

For the $^{136}$Xe $0\nu\beta\beta$ search, dual-phase LXe TPC experiments hold large LXe mass $M$, but are also subject to high $B$ compared to dedicated $0\nu\beta\beta$ experiments. In addition to the dominant material-induced backgrounds, key internal and irreducible sources include the daughter nuclei of $^{222}$Rn, $2\nu\beta\beta$ decay of $^{136}$Xe, $\beta$-decay of $^{137}$Xe, and neutrino-electron scattering of $^8$B solar neutrinos. $^{222}$Rn emanates from the detector materials, with an activity concentration measured to be \SI{0.9}{\micro\becquerel\per\kilogram}~\cite{XENON:2025nic}. This background is mostly $^{214}$Bi, which includes both the component in the TPC and in the LXe shell outside of the TPC. The former includes both the $\beta$-decay and the $\gamma$-ray, while the latter only has the $\gamma$ part. The $2\nu\beta\beta$ decay of $^{136}$Xe, which is naturally occurring in LXe and a non-negligible background in the $0\nu\beta\beta$ search, has a measured abundance of $(8.97\pm0.16)\times10^{-2}$ mol/mol. With a half-life of 3.82 min and a Q-value of 4.17 MeV, the $\beta$-decay of $^{137}$Xe is a relevant background source produced through neutron capture on $^{136}$Xe, and the rate is estimated to be $(6\pm5)$ t$^{-1}$yr$^{-1}$ \cite{XENON:2022evz}. Both the $2\nu\beta\beta$ decay of $^{136}$Xe and solar neutrino-electron scattering backgrounds exhibit highly similar event topologies to that expected from $0\nu\beta\beta$, which makes them irreducible.

\begin{figure}[h]
\centering
\includegraphics[width=0.5\textwidth]{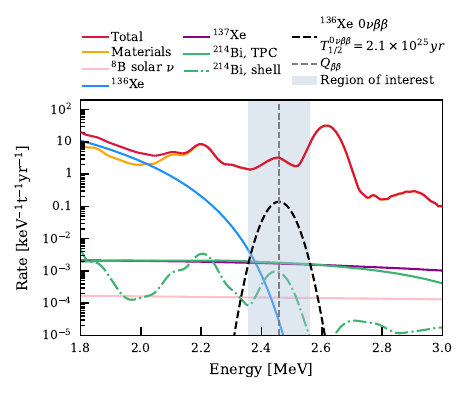}
\caption{Energy spectrum of all backgrounds relevant to the $0\nu\beta\beta$ search in XENONnT. Dominant contributions around Q$_{\beta\beta}$ arise from material background (solid orange) and, in particular from $^{214}$Bi in the TPC (solid green), $^{214}$Bi in the LXe shell (dash green), and $^{137}$Xe (solid purple). $2\nu\beta\beta$ (solid blue) of $^{136}$Xe and $^{8}$B solar neutrinos (solid pink) are subdominant. The shaded light blue area denotes the 2 $\Delta E$ ROI.}
\label{figure-sig_bkg_spectrums}
\end{figure}

The material background constitutes the dominant contribution in the $0\nu\beta\beta$ region of interest. This background originates from the naturally abundant radioactive isotopes within detector materials, namely $^{60}$Co as well as the decay chains $^{238}$U and $^{232}$Th. Figure~\ref{figure-sig_bkg_spectrums} shows the preliminary signal and background spectra of the $0\nu\beta\beta$ search, where it is clear that the main obstacle that limits $0\nu\beta\beta$ sensitivity is the material background. Specifically, there is a gamma peak at 2447 keV originating from $^{214}$Bi, which resides exactly upon the expected 2458\,keV $0\nu\beta\beta$ peak. The presence of this peak significantly degrades the sensitivity of the $0\nu\beta\beta$ search, where the discrimination provided by the A-CNN model becomes particularly valuable.

Material backgrounds are usually mitigated by extensive instrumentation. For example, the \textsc{Majorana Demonstrator} experiment utilized ultra-clean, underground electroformed copper~\cite{MAJORANA:2017qwe}; the EXO-200 experiment employed ultra-low-background copper~\cite{Auger:2012gs}; and the GERDA experiment implemented a LAr active veto system to achieve the world's lowest background rates~\cite{Biancacci:2022etv}. These hardware implementations require significant resources and long-term commitment. However, given that dual-phase LXe TPCs are primarily designed for WIMP dark matter detection, their background optimizations are mainly for low energies rather than the $0\nu\beta\beta$ energy region. Therefore, machine learning remains the most feasible and cost-effective approach for mitigating these backgrounds without substantial hardware and instrumentation investments.

\section{Training Preparation}\label{sec3}

\subsection{MC Simulation}\label{sec31}

The training data are generated by full-chain Monte-Carlo (MC), which uses GEANT4~\cite{GEANT4:2002zbu} simulation as input and provides observable quanta processed by the XENONnT waveform simulator (WFSim)~\cite{wfsim}. Firstly, specific energetic particles are simulated with the entire XENONnT detector geometry by GEANT4, and the simulation outputs are used to calculate the expected produced quanta (photons in the S1 and electrons in the S2). WFSim takes the produced quanta and the position information to simulate the per-PMT response, which makes both high-level features and low-level data such as hit patterns and waveforms available. The simulated S2$_{\text{top}}$ waveforms are used for the A-CNN model training.

The major difference between the $0\nu\beta\beta$ signal and the material background is the nature of SS vs. MS: $0\nu\beta\beta$ signals are mostly SS events that deposit energy in a small volume, while MeV gamma-rays mostly result in MS events that scatter multiple times, leading to multiple distinct energy depositions in the detector. When the depositions are spatially well-separated, the event can be effectively rejected during analysis. However, if the depositions occur at short spatial separations, it may evade background rejection. The machine learning model is trained to accept the 1D waveform of each event and output an SS vs. MS classification score. The training data includes simulated electrons as the SS signal and gammas as the MS background. Electrons and gammas are uniformly generated within the detector; the energies of the simulated events are also uniformly distributed between 0 and 4 MeV.

\begin{figure}[h]
\centering
\includegraphics[width=0.45\textwidth]{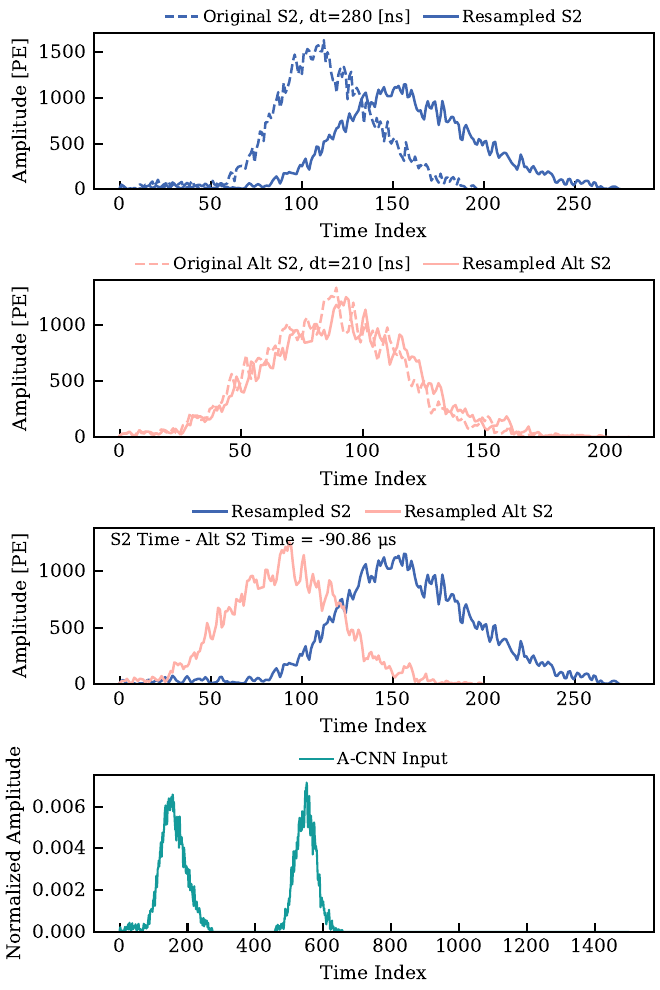}
\caption{Training input construction example. First (from top to bottom): the main S2$_{\text{top}}$ and its resampled waveform; Second: the alternative S2$_{\text{top}}$ and its resampled waveform; Third: the aligned resampled main S2$_{\text{top}}$ and alternative S2$_{\text{top}}$ comparison; Fourth: the complete A-CNN input waveform.}
\label{figure-resampled_waveform}
\end{figure}

The next step is to process the simulated 1D waveforms into time series format. In the current data structure, regardless of how physically wide the S2s are in real time, they are stored as arrays of 200 samples. Therefore, the sampling time $dt$ is different for different S2s. In cases where multiple S2s are present in the same event, the waveforms of the main S2$_{\text{top}}$ (the top PMT waveform of the largest S2 in the event) and the alternative S2$_{\text{top}}$ (the second largest) are first resampled to a constant physical sampling time $dt=200$ ns. These waveforms are then concatenated in temporal order and normalized to unit area, resulting in unified inputs for model training. Figure~\ref{figure-resampled_waveform} illustrates an example of the process. In the event, the main S2$_{\text{top}}$ and the alternative S2$_{\text{top}}$ are originally stored in a sampling time $dt$ of 270 ns and 210 ns respectively. They are both resampled to $dt=200$ ns first, as shown in Figure~\ref{figure-resampled_waveform} (First and Second). The resampled main S2$_{\text{top}}$ and alternative S2$_{\text{top}}$ are aligned and compared in Figure~\ref{figure-resampled_waveform} (Third), and their time difference is \SI{90.86}{\micro\second}. Then this time difference is used to concatenate the alternative S2$_{\text{top}}$ and the main S2$_{\text{top}}$ into a one-dimensional time series containing both peaks, as shown in Figure~\ref{figure-resampled_waveform} (Fourth). The variance of the concatenated lengths is then captured by filling the remaining samples with zeros in the final fixed-length waveforms of 1500 samples, for model training purpose. Lastly, the final time series data are normalized to have an area of one.

\subsection{Augmentation}\label{sec32}

As the model trained on simulated events is intended for deployment in real data analysis, ensuring robustness and generalization is essential. The two primary considerations are outlined below:
\begin{itemize}
    \item \textbf{S2 Width:} the width of simulated S2 waveforms (e.g., the \emph{50\% area width} $w_{50}=t_{75}-t_{25}$, where $t_{p}$ denotes the time at which the cumulative S2 area reaches $p\%$ of its total). Since S2 width is known to be a key parameter for SS vs. MS classification, data augmentation was applied to improve model robustness.
    \item \textbf{Energy Bias:} Machine learning models can be biased when the training data follow a different energy distribution. Although the time series data have been normalized to a unit area, energy bias could still linger within the event baseline, because higher S2 could lead to a more suppressed baseline during the normalization process.
\end{itemize}

To ensure better robustness of the model, we implemented a data augmentation technique~\cite{2017arXiv171204621P} during the machine learning model's training phase. Effective augmentation techniques include cropping, resampling, noise injection, masking, and various other approaches~\cite{Shorten2019AugSurvey}. By training models to produce consistent outputs under varying forms of data augmentation, we developed a model that demonstrates resilience to changes introduced by the applied augmentation methods.

In this work, we adopted two specific data augmentation methods to address the two major causes of discrepancy. The first augmentation is random upsampling: for each training event, we randomly stretch the width of the simulated S2$_\text{top}$ waveform by a certain percentage. The model trained on these randomly upsampled waveforms will be robust against the inconsistent S2 widths between MC simulations and real data. The second augmentation is random noise, where Gaussian noise randomly sampled from a normal distribution is added to each time index of the time-series input data. The randomly added noise can overshadow the baseline to create a model that is robust against energy bias. The strength of the augmentations is controlled to ensure they do not degrade the model's performance.

With the two augmentation techniques described above, the model achieves strong adaptability and performs well across a wide energy range, achieving excellent consistency between full-chain MC and real data, as demonstrated in Section~\ref{sec42}.

\subsection{Model Structure}\label{sec33}

The machine learning model employs a CNN architecture which includes two key components: a feature extractor and a linear classifier. The feature extractor is designed to progressively capture and refine the time-series patterns inherent in the input data, and the linear classifier combines the patterns into a single prediction. As detailed in Section~\ref{sec31}, the input for each event is generated by concatenating the main S2$_{\text{top}}$ and alternative S2$_{\text{top}}$ waveforms according to their order of occurrence and relative time differences, resulting in a single-channel time series of length 1500. The model ultimately outputs a probability between 0 and 1, representing the likelihood that an event is signal-like (1) or a background-like (0).

\begin{figure*}[htbp]
\centering
\includegraphics[width=0.92\textwidth]{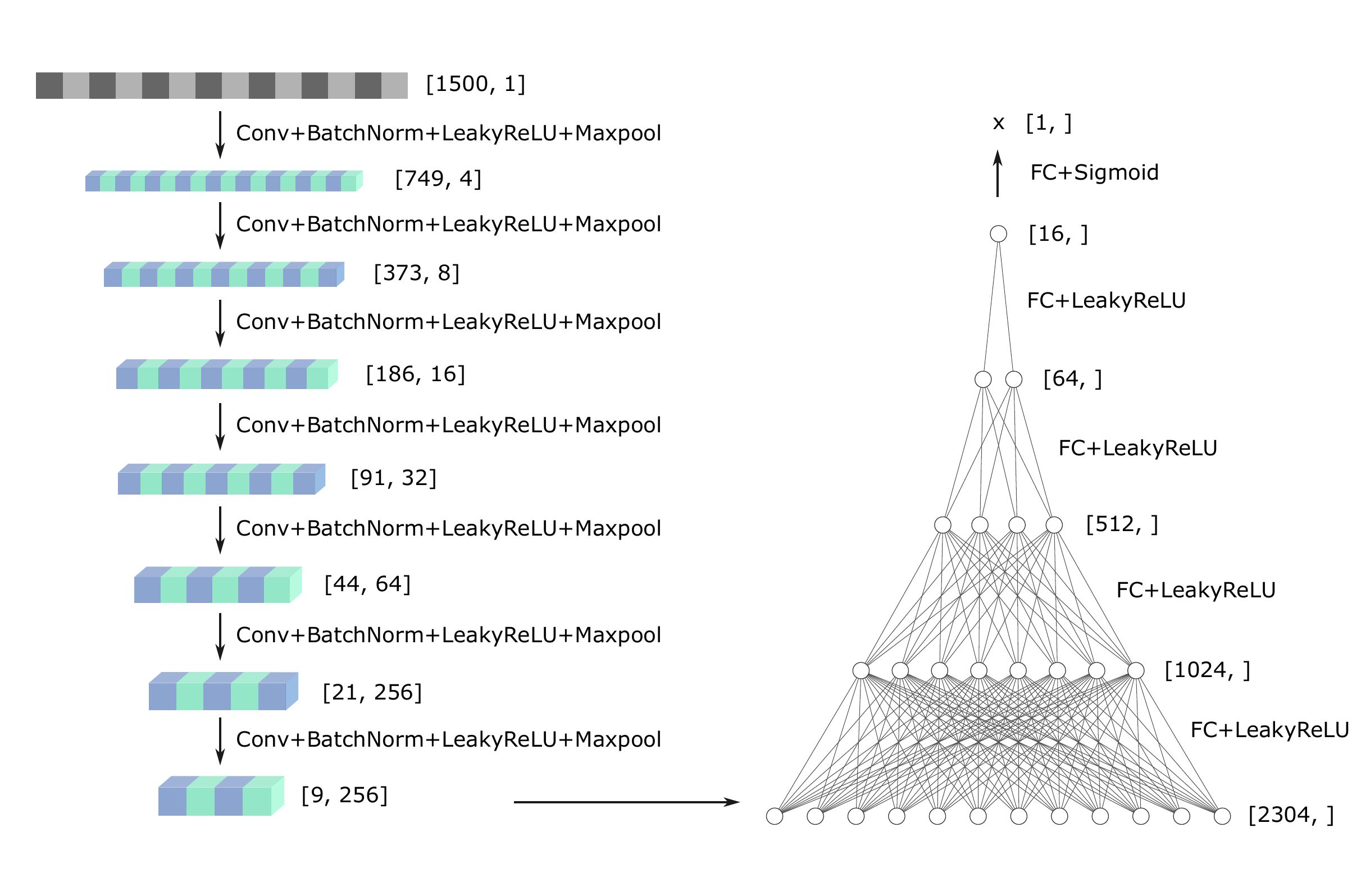}
\caption{Schematic diagram of the A-CNN model. Left: the feature extractor; Right: the linear classifier.}
\label{figure-ACNN_structure}
\end{figure*}

Figure~\ref{figure-ACNN_structure} (Left) illustrates the detailed network structure of the feature extractor. Starting with the one-dimensional time-series, the convolutional layers~\cite{lecun1998gradient} extract pattern information with LeakyReLU activation functions~\cite{maas2013rectifier} and expand the representation into multiple channels, while max-pooling layers~\cite{lecun1998gradient} control the tensor dimensions without sacrificing essential details. Batch normalization~\cite{ioffe2015batch} is incorporated to accelerate convergence during training. Through this series of operations, the data are transformed into a tensor of shape [9, 256].

Following the feature extraction stage, the flattened tensor (length 2304) is fed into the fully connected classifier, as shown in Figure~\ref{figure-ACNN_structure} (Right). At this stage, the dimensionality of the data is progressively reduced through successive linear layers, effectively combining the extracted features for classification. A final sigmoid function maps the output onto an A-CNN score between 0 and 1, which is then used in subsequent classification tasks and $0\nu\beta\beta$ studies.

\section{Training Results}\label{sec4}

\subsection{Classification Power}\label{sec41}

Using the full-chain simulation introduced in Section~\ref{sec31}, a total of 40,000 electron ($0\nu\beta\beta$ signal-like) events and 40,000 gamma-ray (background-like) events, homogeneously distributed within the TPC, were simulated, whose energies were uniformly distributed between 0 and 4 MeV. These events were subsequently divided into training and validation sets with a 3:1 ratio. An independent test dataset consisting of 20,000 electron and 20,000 gamma events was also simulated to evaluate the model performance. All time-series data were augmented using the method described in Section~\ref{sec32} before being fed into the convolutional layers.

\begin{figure*}[!t]
\centering
\includegraphics[width=0.9\textwidth]{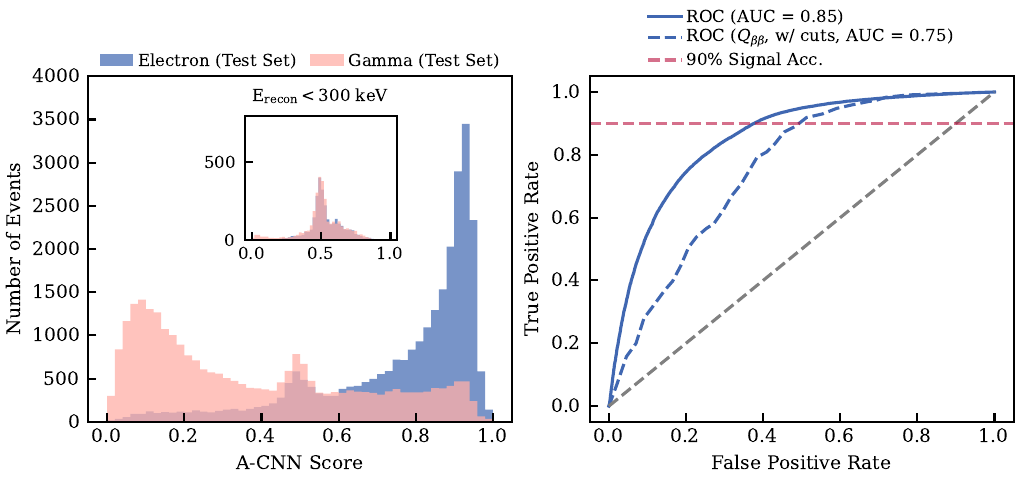}
\caption{The trained A-CNN performance on the test set. Left: A-CNN score distribution on 0-4 MeV events. The inset shows events with energy lower than 300 keV, which the population at A-CNN score around 0.5 mostly comes from low-energy events. Right: the A-CNN ROC curve without traditional cuts (solid blue, 62\% background rejection at 90\% signal acceptance), the ROC curve with traditional cuts in $Q_{\beta\beta}$ (dashed blue, 50\% background rejection at 90\% signal acceptance), and the 90\% signal acceptance (TPR) line (dashed red). Background rejection = 1-FPR.}
\label{figure-roc}
\end{figure*}

The optimization objective (loss function) used in A-CNN training is the binary cross-entropy (BCE) loss \cite{cox1958regression}.
\begin{equation}
\mathcal{L}(\mathbf{x}, \mathbf{y}) = - \frac{1}{N} \sum_{i=1}^{N} \left[ y_i \cdot \log(x_i) + (1 - y_i) \cdot \log(1 - x_i) \right],
\label{eq:loss}
\end{equation}
where \(x_i\in(0,1)\) denotes the predicted probability for the positive class of sample \(i\) and \(y_i\in\{0,1\}\) denotes the corresponding ground-truth label. This expression represents the BCE averaged over a mini-batch of size \(N\); it penalizes disagreement between predicted probabilities and binary targets: the term \(y_i\log x_i\) contributes when \(y_i=1\), whereas \((1-y_i)\log(1-x_i)\) contributes when \(y_i=0\).
Minimizing \(\mathcal{L}\) drives the predictions toward the true labels. Training employs the Adam optimizer~\cite{kingma2014adam} and typically converges in about 20 epochs, with close agreement between training and validation metrics.

Figure~\ref{figure-roc} (Left) shows the performance of the A-CNN model on the test set. A clear separation is observed between the A-CNN score distribution of signal and background events, albeit a small overlapping peak around 0.5. As shown in the embedded figure, this peak primarily contains events with energies below 300 keV, far outside the $0\nu\beta\beta$ region of interest. These low-energy events are inseparable because they contain much fewer ionization electrons, which play a crucial role in SS/MS separation. Notably, the A-CNN model assigns a score of 0.5 to these low-energy events, reflecting the model's uncertainty in classifying them as either SS or MS. The approximately symmetric accumulation of near-\(0.5\) scores from both classes suggests comparable fractions of hard-to-classify events in the training set, consistent with effective dataset construction and augmentation.

The receiver operating characteristic (ROC) curve for the A-CNN model is shown in Figure~\ref{figure-roc} (Right), which has an area under the curve (AUC) of 0.85. The x-axis represents the false positive rate (FPR), defined as the proportion of background events incorrectly classified as signal. It is the complement of the background rejection rate, i.e., FPR $ = 1 - $ background rejection rate. The y-axis represents the true positive rate (TPR), defined as the proportion of actual positive cases that are correctly identified by the model, is equivalent to signal acceptance in the $0\nu\beta\beta$ search. If a 90\% signal acceptance is chosen, the corresponding background rejection rate is 62\%. Another evaluation is performed by applying the data quality cuts and S2 SS cut, a traditional approach for rejecting MS events by removing events with a large ratio between the alternative S2 area and the main S2 area~\cite{XENON1T_data_selection}. After applying these traditional cuts, the A-CNN model is still able to reject 50\% of the remaining background events in the Q$_{\beta\beta}$ region, see Figure~\ref{figure-roc} (Right).

\subsection{Data-MC Validation}\label{sec42}

Although the trained A-CNN model performs well on the simulated test dataset, it is critical to ensure that the model provides the same response between simulated events and real detector events. This data-MC comparison is necessary to ensure consistent model performance between data and MC, which would demonstrate that the model can be used for physics analysis. The study was performed over three different datasets including both simulated waveforms and real detector data. These dataset selections allow us to verify model consistency across different types of events and a wide energy range:
\begin{enumerate}
    \item The first dataset includes $^{136}$Xe $2\nu\beta\beta$ decay events. To select a relatively pure $2\nu\beta\beta$ dataset, events of energy within [980, 1080] keV and within a central 1 t fiducial volume (FV) were selected. With this energy range, the contamination of $^{214}$Pb events is suppressed, and the leakage of material gamma-rays is also small. To maintain consistency and match realistic application conditions, both the TPC data and the corresponding MC simulation passed the data cleaning cuts and the S2 SS cut. 
    \item The second dataset includes background-like events from $^{208}$Tl gamma-rays. The energy of the gamma peak lies at 2614.5 keV so that events with energy within [2550, 2650] keV and within a 3 t FV are selected. The reason for a larger FV as opposed to the signal-like events is to ensure sufficient statistics for TPC data, as the material gamma-ray rate decays exponentially as the isotropic distance from any walls in the TPC increases. The same set of cuts was applied. 
    \item The third dataset includes $\beta$-decays followed by a $\gamma$ emission from $^{212}$Pb events. These events may also be distinguished by the A-CNN model when the time interval between the $e^-$ and $\gamma$ emission is short. Here $^{212}$Pb events were selected from TPC data with an energy cut of [300, 500] keV in a 3 t FV. The same cuts were also applied for both data and MC. 
\end{enumerate}

\begin{figure}[hbtp]
    \centering
    \begin{minipage}{0.43\textwidth}
        \centering
        \includegraphics[width=\textwidth]{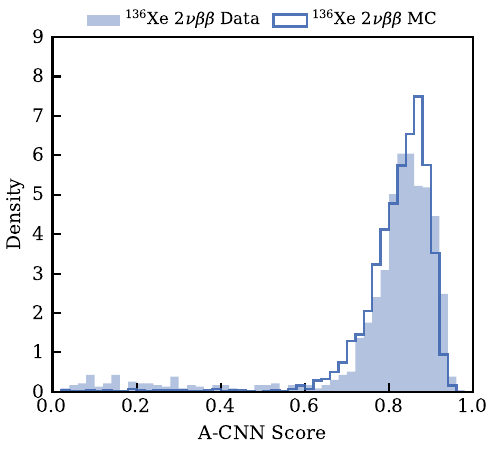}
    \end{minipage}%
    \hspace{0.02\textwidth} % Optional space between images
    \begin{minipage}{0.43\textwidth}
        \centering
        \includegraphics[width=\textwidth]{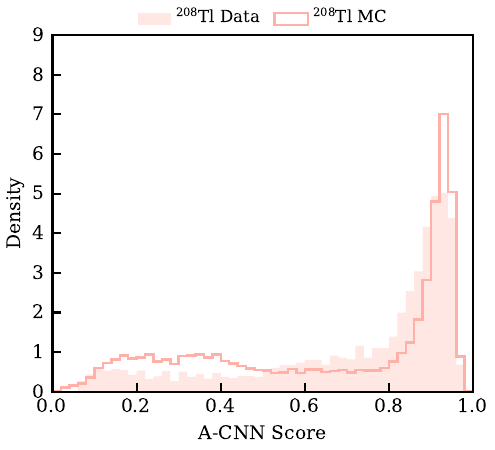}
    \end{minipage}%
    \hspace{0.02\textwidth} % Optional space between images
    \begin{minipage}{0.43\textwidth}
        \centering
        \includegraphics[width=\textwidth]{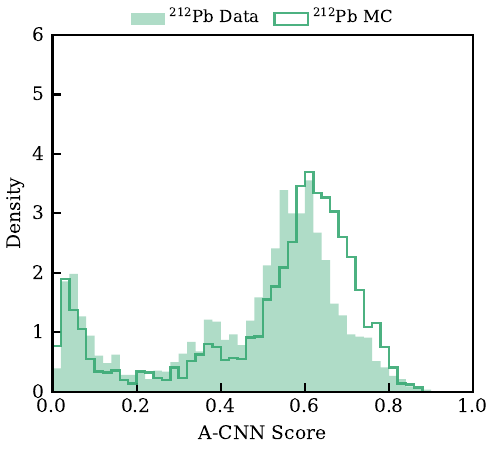}
    \end{minipage}
    \caption{Data-MC validation of the A-CNN model through signal vs. background scoring after the traditional cuts. Top: A-CNN score spectrum for MC/data of signal-like events ($^{136}$Xe $2\nu\beta\beta$); Middle: background-like events ($^{208}$Tl gamma-peak); Bottom: $\beta+\gamma$ events ($^{212}$Pb).}
    \label{figure-data_MC}
\end{figure}

Figure~\ref{figure-data_MC} (Top) shows the result of the data-MC validation study for signal-like events. We observe overall consistency between data and simulation in the A-CNN score distributions, with residual differences at the few-percent level discussed below. Figure~\ref{figure-data_MC} (Middle) demonstrates that the data-MC agreement is also satisfactory for background-like events. Notably, these events exhibit a peak at high A-CNN scores, as traditional cuts have already removed the more easily identifiable background events, leaving only the more signal-like residuals. Figure~\ref{figure-data_MC} (Bottom) shows the A-CNN score spectra for $\beta+\gamma$ events. Although the model was not explicitly trained on this type of event, it successfully distinguishes $\beta+\gamma$ events from other types and demonstrates good agreement between full-chain MC and real data. This may be useful for other physics searches where $^{212}$Pb and $^{214}$Pb contribute a considerable fraction of the background.

The consistent agreement between data and simulation across different event types and energies supports the application of the A-CNN model to the $0\nu\beta\beta$ search in dual-phase LXe TPCs. The A-CNN score is expected to serve as an additional cut on top of the traditional cuts to improve signal and background classification, and the score threshold can be optimized for the best $0\nu\beta\beta$ sensitivity. There is a discrepancy of up to 3\% between the event acceptance of data and MC, and this discrepancy is accounted for as a systematic uncertainty in the efficiency of  $0\nu\beta\beta$ decay search. As a comparison, the EXO-200 experiment~\cite{EXO-200:2019rkq} reported a data-MC discrepancy systematic uncertainty of about 10\%, while the KamNet in KamLAND-Zen experiment~\cite{Li:2022frp} reported a data-MC discrepancy systematic uncertainty of about 3\% and was used as an event selection cut. The data-MC agreement achieved by the A-CNN model is notably superior to both models over a large energy range. This confirms the robustness of the A-CNN model with data augmentation, which effectively captures the underlying data distributions and nuances that are often missed by ML models.

\subsection{Interpretability Study}\label{sec43}

\begin{figure}[H]
\centering
\includegraphics[width=0.50\textwidth]{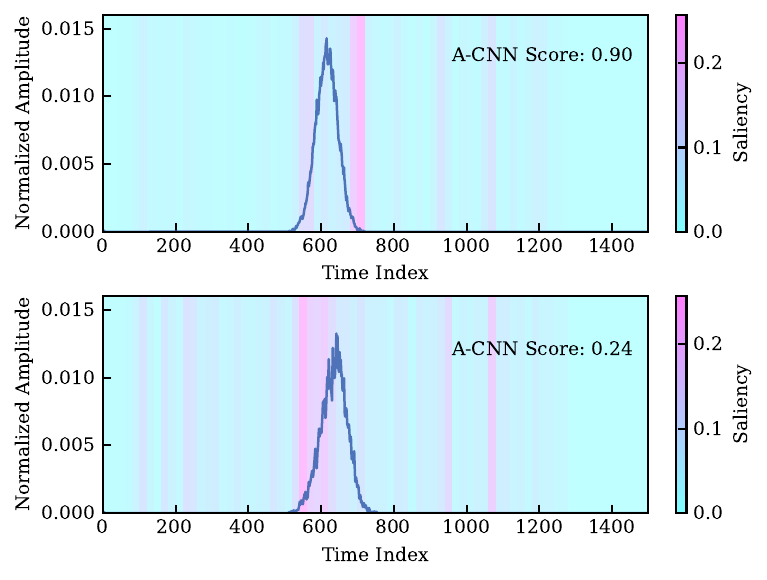}
\caption{The saliency plots for the A-CNN. The shape of the waveforms shows the time profile of the XENONnT events, and the color shows the relative saliency score of each time slice.}
\label{figure-interpretability}
\end{figure}

To understand the classification capabilities of the trained A-CNN model, we conducted an interpretability study using saliency maps~\cite{Simonyan14a}, a method that has been widely adopted by the artificial intelligence and machine learning (AI/ML) community. These maps were generated by calculating the gradient of the model's output with respect to the input waveform. Following standard backpropagation~\cite{Rumelhart1986}, we computed the saliency by gradient propagation through the network, implemented with PyTorch autograd~\cite{Paszke2019}. The calculated gradient quantifies how incremental changes in various segments of the waveform influence the model's predictions. Regions exhibiting larger gradients correspond to areas that are more important in the decision-making process. The gradient and corresponding waveform can be visualized using saliency plots, as shown in Figure~\ref{figure-interpretability}.

To demonstrate the interpretability of the A-CNN model, we selected one representative signal and one representative background waveform by the trained A-CNN model. Figure~\ref{figure-interpretability} shows a typical waveform of high A-CNN score (0.90) at the top and a typical event waveform of low A-CNN score (0.24). The main difference between the two waveforms is that the signal-like waveform with high A-CNN score has fewer fluctuations, while the background-like waveform with a low score has more fluctuations. The saliency map shows that the trained A-CNN pays attention to those fluctuations to perform signal vs. background classification.  

It is also worth noting that the saliency map shows greater attention to the rising and falling edges of the waveform. When the main S2 peak results from merged multi-scatter events, the larger scatter amplitudes are typically less affected. However, the smaller scatters may alter the shape of the waveform either before or after the main peak, depending on the order of occurrence and the relative distance between the scatters. The A-CNN model may be affected by baseline noise that is omnipresent throughout the waveform. This effect is exacerbated if the waveform has low amplitude relative to the baseline noise. However, as described in Section~\ref{sec32}, the random noise data augmentation techniques can efficiently mitigate the bias arising from it. In summary, the saliency maps suggest that the model focuses primarily on structured features, such as the rising and falling edges of the S2 peak, rather than random noise.

\section{\texorpdfstring{$0\nu\beta\beta$}{0nuBB} Sensitivity Boost}\label{sec5}

In the $0\nu\beta\beta$ energy ROI, the A-CNN model was able to reject 50\% of the material background while keeping 90\% of the $0\nu\beta\beta$ signal even after the traditional cuts were applied. This discrimination power can be translated into a sensitivity boost through two dedicated sensitivity studies described in this section.

The first sensitivity study used a counting experiment setup. We first counted the number of events within the $0\nu\beta\beta$ energy ROI and compared it against the expected background in the same energy region. This method usually yields a conservative estimate of the sensitivity since it does not take into account the spectral shape of signal and background. In this model, XENONnT's sensitivity to $0\nu\beta\beta$ decay is approximated as proportional to $S/\sqrt{B}$~\cite{Li:2022frp}, where $S$ is the number of signal events and $B$ is the number of background events in the ROI. Since the background is dominated by material gamma-rays in the ROI, and the A-CNN classifier rejects 50\% of the background while preserving 90\% of the signal, the sensitivity was boosted by

\begin{equation}
T_{1/2}^{0\nu\beta\beta} \propto \frac{90\%S}{\sqrt{(100\%-50\%)B}} = 1.273 \times \frac{S}{\sqrt{B}},
\label{eq:sens_boost}
\end{equation}

\noindent where 90\% is the TPR and one minus 50\% (true negative rate, TNR) is the FPR, which indicates the percentage of backgrounds remaining after the classification. The number 1.273 corresponds to a 27.3\% boost in $0\nu\beta\beta$ sensitivity.

\begin{figure}[H]
\centering
\includegraphics[width=0.5\textwidth]{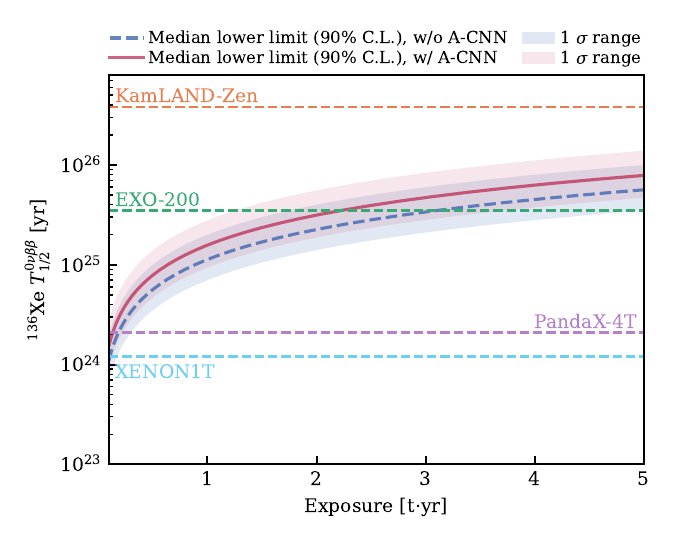}
\caption{Projected $^{136}$Xe $0\nu\beta\beta$ sensitivity of XENONnT at different exposures without A-CNN (solid red) and with A-CNN (dashed blue), and comparison with the results of XENON1T (cyan), PandaX-4T (purple), EXO-200 (green) and KamLAND-Zen (orange).}\label{figure-sens_boost}
\end{figure}

In the second sensitivity study, we performed a likelihood fit to the signal and background energy spectra to obtain a better estimate of the sensitivity improvement. We leveraged the background models described in Section~\ref{sec2} and ran toy MCs to estimate the sensitivity of XENONnT science data. The energy resolution derived from the first and second science run data was utilized in this study. A preliminary fiducial volume was selected and spectrum fits were then conducted with the A-CNN cut systematics considered. For each toy MC, we used the best-fit $0\nu\beta\beta$ signal rate and Wald's approximation~\cite{Cowan:2010js} to find the upper limit of $0\nu\beta\beta$ rate, which was then converted to the lower limit of $^{136}$Xe $0\nu\beta\beta$ decay half-life. Eight thousand toy MCs were simulated and the median lower limits of the $0\nu\beta\beta$ half-lives are calculated for models without and with the A-CNN. The projected sensitivity boost from the A-CNN classifier was 39.4\%. The result is shown in Figure~\ref{figure-sens_boost} and compared to other experiments. Note that Eq.~\ref{eq:halflife-scaling} also indicates that the $0\nu\beta\beta$ sensitivity scales inversely with the energy resolution $\Delta E$; therefore, further sensitivity gains can be expected if the energy resolution improves.

This sensitivity gain demonstrates that the A-CNN improves XENONnT’s sensitivity to $0\nu\beta\beta$ decay. Further improvements are plausible with models operating on lower-level spatiotemporal waveforms, which retain per-channel timing and topology information that is partially lost in higher-level summaries.

\section{Conclusion and Outlook}\label{sec6}

We developed an A-CNN model to address the material background limitations for the $0\nu\beta\beta$ search of dual-phase LXe TPC experiments, a major constraint on experimental sensitivity. While these backgrounds traditionally require substantial hardware investments in detector purification and veto systems, our machine-learning-based approach achieves significant background rejection without any hardware upgrade. The trained A-CNN demonstrates powerful suppression of the material background, resulting in an improvement of about 40\% in the projected experimental sensitivity demonstrated in XENONnT. This gain is enabled by using low-level waveforms, which preserves more event information beyond traditional analysis. It is further supported by a targeted data-augmentation scheme that mitigates residual data–MC discrepancies and improves generalization. Through extensive validation, our model demonstrated good agreement between Monte Carlo simulations and experimental data, with interpretability studies providing insight into the model's classification mechanisms.

In the future, we plan to apply the A-CNN on the XENONnT data to improve the search sensitivity for $0\nu\beta\beta$ as well as the next-generation dual-phase LXe TPC experiments such as XLZD~\cite{Baudis:2024jnk, XLZD:2024pdv}. With 60-80 t of natural xenon in the sensitive target, our model provides a cost-effective approach for XLZD to achieve improved $0\nu\beta\beta$ sensitivity without compromising its primary WIMP search mission or requiring additional hardware investments. Recently, SS and MS classification based on hit patterns has been explored~\cite{Sazzad:2025own}. As we look ahead, we plan to explore more sophisticated architectures, including spatiotemporal deep learning models, to further enhance background rejection capabilities and expand the potential of software-based solutions in rare-event searches.

\section{Acknowledgment}\label{sec7}

We gratefully acknowledge support from the National Science Foundation, Swiss National Science Foundation, German Ministry for Education and Research, Max Planck Gesellschaft, Deutsche Forschungsgemeinschaft, Helmholtz Association, Dutch Research Council (NWO), Fundacao para a Ciencia e Tecnologia, Weizmann Institute of Science, Binational Science Foundation, Région des Pays de la Loire, Knut and Alice Wallenberg Foundation, Kavli Foundation, JSPS Kakenhi, JST FOREST Program, and ERAN in Japan, Tsinghua University Initiative Scientific Research Program, National Natural Science Foundation of China, Ministry of Education of China, DIM-ACAV+ Région Ile-de-France, and Istituto Nazionale di Fisica Nucleare. This project has received funding/support from the European Union’s Horizon 2020 and Horizon Europe research and innovation programs under the Marie Skłodowska-Curie grant agreements No 860881-HIDDeN and No 101081465-AUFRANDE.

We gratefully acknowledge support for providing computing and data-processing resources of the Open Science Pool and the European Grid Initiative, at the following computing centers: the CNRS/IN2P3 (Lyon - France), the Dutch national e-infrastructure with the support of SURF Cooperative, the Nikhef Data-Processing Facility (Amsterdam - Netherlands), the INFN-CNAF (Bologna - Italy), the San Diego Supercomputer Center (San Diego - USA) and the Enrico Fermi Institute (Chicago - USA). We acknowledge the support of the Research Computing Center (RCC) at The University of Chicago for providing computing resources for data analysis.

We thank the INFN Laboratori Nazionali del Gran Sasso for hosting and supporting the XENON project.

\bibliography{main}% Produces the bibliography via BibTeX.

\end{document}